**Charge Transfer in Ternary Solar Cells Employing Two Fullerene Derivatives: Where do Electrons Go?**

Dr. Andreas Sperlich[a]*, Dr. Michael Auth[a], Prof. Dr. Vladimir Dyakonov[a]*
[a]Experimental Physics 6, Julius Maximilian University of Würzburg, Würzburg, Germany
*sperlich@physik.uni-wuerzburg.de, dyakonov@physik.uni-wuerzburg.de

**Abstract**
Earlier reports demonstrated that ternary organic solar cells (OSC) made of donor polymers (D) blended with different mixtures of fullerene acceptors (A:A) performed very similarly. This finding is surprising, as the corresponding fullerene LUMO levels are slightly different, which might result in decisive differences in the charge transfer step. We investigate ternary OSC (D:A:A) made of the donor polymer P3HT with stoichiometric mixtures of different fullerene derivatives, $PC_{60}BM:PC_{70}BM$ and $PC_{70}BM:IC_{60}BA$, respectively. Using quantitative electron paramagnetic resonance (EPR) we can distinguish 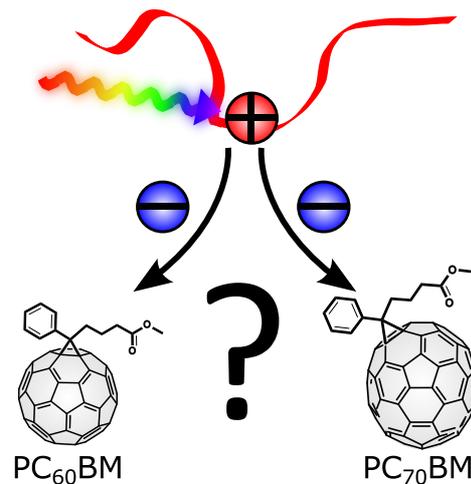 between positive and negative polarons, localized on the specific molecules. We found that after the initial charge transfer step, the electrons are re-distributed over two nearby acceptors in agreement with their stoichiometry and their relative LUMO energy difference. Remarkably, the measured ΔLUMO differences in fullerene mixtures are reduced by an order of magnitude compared to that of the pristine materials, i.e., below 1 meV for $PC_{60}BM:PC_{70}BM$ and $(20 \pm 5)$ meV for $PC_{70}BM:IC_{60}BA$. Furthermore, we found that this reduced ΔLUMO explains the shift in open circuit voltage for D:A:A organic solar cells. We attribute these findings to hybridization, leading to an effective fullerene LUMO. Consequently, multi-acceptor blends are indeed a viable option for photodetectors and solar cells, as they combine the best electron acceptor and light absorbing properties.

## 1. Introduction

Organic photovoltaic (OPV) cells employing fullerene derivatives as acceptors (A) have high potential for direct sun to electricity conversion with reported power conversion efficiencies above 11 %.[1,2] Thereby, the highest-performing devices use $C_{70}$- instead of $C_{60}$-derivatives as electron acceptor.[3] This is because $C_{70}$ exhibits a stronger optical absorption in the visible part of the solar spectrum, which helps to convert more of the incident photons into electricity.[4] High-performing donor (D) materials, like donor-acceptor copolymers are usually very sensitive for blue and red photons. However, they lack absorption just in the spectral range where fullerenes can enhance the absorbance of the photo active layer. The commonly used material is the soluble $C_{70}$ derivative $PC_{70}BM$, but it is prohibitively expensive for material purified to >99.5%. Yet, the most common impurity in $C_{70}$-based fullerenes is again $C_{60}$. To see if purification to this extent is necessary, we investigated the OPV performance of "unpurified" fullerene materials, i.e., intentionally prepared fullerene mixtures. For ternary (D:A:A) bulk heterojunction (BHJ) solar cells we observed a gradual shift of the open circuit voltage ($V_{OC}$) upon changing the fullerene stoichiometry. This unexpected $V_{OC}$ shift made us curious to reveal the physical processes taking place on the charge carrier level. Recent studies reported similar tuning effects, regarding the ionization energies and bandgaps of organic semiconductors by blending different derivatives of the same base molecules.[5,6,7,8,9] Therefore, we focused in particular on the effect of mixing fullerenes on the molecular energy levels, since these have a crucial influence on the voltage of the OPV device.

To get an overview of the involved energetics, we compare the highest occupied molecular orbital (HOMO) and lowest unoccupied molecular orbital (LUMO) levels of the fullerene derivatives $PC_{60}BM$, $PC_{70}BM$ and $IC_{60}BA$ in **Figure 1**. A direct comparison of cyclic voltammetry of $PC_{60}BM$ and $PC_{70}BM$ shows that the HOMO/LUMO levels of $PC_{60}BM$ and $PC_{70}BM$ differ by about 6 – 10 meV.[10,11,12,13] On the other hand, a study on fullerene-based diodes by Schafferhans et al. has shown that despite similar electronic trap densities in both $PC_{60}BM$ and $PC_{70}BM$ diodes, the traps are energetically deeper in devices employing $PC_{70}BM$.[14,15] A larger difference is observed for $IC_{60}BA$ and $PC_{70}BM$, where LUMO levels are shifted by 170 – 210 meV.[16,10,17,18,19,20] Even more so, the LUMO of $PC_{84}BM$ is shifted by as much as ≈300 meV with respect to the other two PCBM-derivatives, resulting in enhanced electron trapping.[21] The deliberate doping of polymer:$PC_{60}BM$ solar cells with small amounts of $PC_{84}BM$ was even used to directly study the deteriorating effect of deep trap states in the fullerene-phase on solar cell performance and recombination mechanisms.[22] Hence, it is mandatory to remove higher fullerenes (>$C_{70}$) in the purification process of the $C_{60}$/$C_{70}$ mixtures.

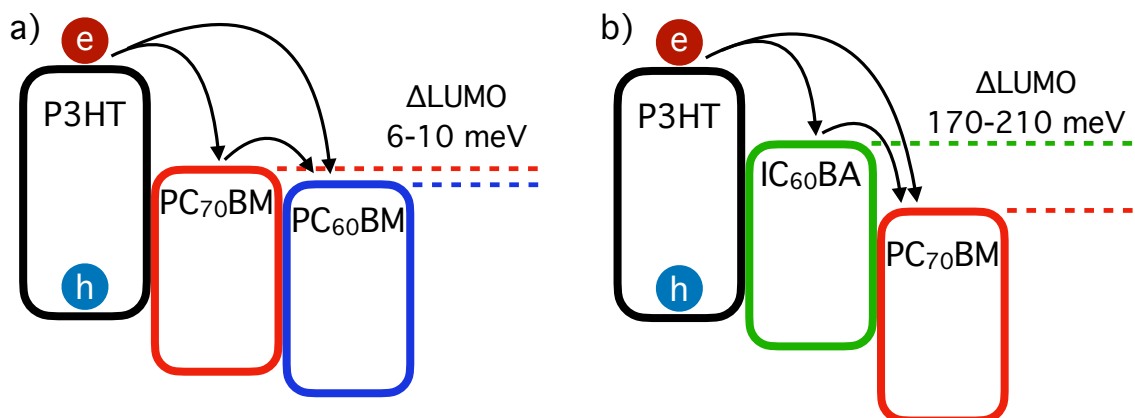

**Figure 1**. HOMO and LUMO levels of pristine materials. The expected fullerene LUMO level differences in a blend would be **a)** 6 – 10 meV for $PC_{70}BM$:$PC_{60}BM$ and **b)** 170 – 210 meV for $PC_{70}BM$:$IC_{60}BA$. The arrows represent possible light-induced charge transfer pathways indicating the question of electron distribution among the available acceptors in blends.

The reported differences in the physical properties of, in particular, $PC_{60}BM$ and $PC_{70}BM$ are small, but nonetheless they show that the molecules are not energetically identical. The question arises, how the energetic differences transfer to actual solar cell devices. If the LUMO levels are indeed different for the studied acceptors, this could lead to enhanced electron trapping on the energetically lower lying fullerene and might deteriorate device performance.

Electron Paramagnetic Resonance (EPR) has a long tradition in OPV research and was actually pivotal in the initial identification of the photo-induced charge transfer in BHJ solar cells.[23] Building on this, we have reported EPR spectra of electrons and holes in blends of conjugated polymers with $C_{60}$- and $C_{70}$-derivatives. We identified the spectral signatures of electrons localized on the different fullerene derivatives which makes them easily distinguishable according to the Landé $g$-factor.[24,25,26] In this study, we utilize the knowledge about the spectroscopic EPR signatures of fullerene radical anions to quantify the spin density and hence the electron/hole distribution in ternary blends of conjugated donor polymers with stoichiometric mixtures of fullerene acceptor derivatives (D:A:A). From the electron distribution, we are able to estimate the actual energetic difference for an electron being localized on either fullerene. This energy difference is the actual ΔLUMO difference in blends of the employed fullerenes, which as we will demonstrate, deviates significantly from the reported literature values for pristine materials. We then show that this behaviour can explain the gradual $V_{OC}$ shift in stoichiometric ternary solar cells.

## 2. Experimental Section

*Materials:* The polymer P3HT (poly(3-hexylthiophene)) was purchased from Aldrich. The low-band-gap donor-acceptor copolymers PCDTBT (a poly(2,7-carbazole) derivative) and PTB7 (consisting of thieno[3,4-b]thiophene and benzodithiophene monomers) were purchased from 1-material. The fullerene-derivatives are from Solenne BV. Known impurities are <0.1 % $C_{60}$ and $C_{70}$, while higher fullerenes (such as $C_{84}$), are excluded to be possible impurities (≪0.1 %). Overall purity of fullerene-derivatives ($PC_{70}BM$, $PC_{60}BM$, $IC_{60}BA$) is >99 %.

*Sample Preparation* took place inside a nitrogen glovebox to avoid exposure to oxygen. The materials were dissolved in chlorobenzene with a concentration of 20 mg/ml and stirred overnight. Polymer and fullerene solutions were mixed in the desired blend ratio. P3HT was blended with equal amounts of fullerenes (1:1), while PCDTBT and PTB7 were blended with fullerenes in ratios of 1:4 and 1:1.5, respectively, as these are approximately the best mixing ratios considering solar cell performance. The fullerenes' stoichiometric weight ratio $X$ was varied from $X = 100$ (pure $PC_{70}BM$) to $X = 0$ (pure $PC_{60}BM$ or $IC_{60}BA$). 0.2 ml portions of the solutions were poured into EPR tubes, vacuum-dried for one hour and sealed under ≈10 mbar helium atmosphere.

*EPR Spectroscopy:* We used a modified X-Band spectrometer (Bruker E300). Samples were placed in a resonant cavity (Bruker ER4104OR) with optical access and a continuous flow helium cryostat insert (Oxford ESR 900). Illumination was provided by a 532 nm laser arriving on the sample with 20 mW. First derivative EPR spectra were recorded with lock-in detection, referenced by a magnetic field modulation amplitude of 0.05 mT. The $g$-factor was calibrated for each measurement with the help of a NMR Gaussmeter (Bruker ER035M) and a microwave frequency counter (EIP 28b). For each temperature, the microwave power was tuned to avoid signal distortions due to saturation effects.

*Solar Cell Production* was done by first spin coating a 40 nm thick layer of PEDOT:PSS (poly(3,4-ethylenedioxythiophene):poly-(styrenesulfonate), Heraeus Clevios AL4083) on top of a glass substrate covered with patterned 150 nm thick indium tin oxide (ITO) electrodes. This layer was thermally annealed for 10 minutes at 130°C. The 100 nm active layer was spin coated inside a nitrogen glove box followed by thermal annealing (10 minutes at 130°C). A calcium (3 nm) / aluminum (120 nm) cathode was deposited in an evaporation chamber (pressure <$10^{-6}$ mbar). A shadow mask results in an active area of 3 $mm^2$. *J-V* characteristics were measured with a SMU (Keithley) under simulated AM1.5G illumination.

## 3. Ternary Solar Cell Performance

**Figure 2** shows the J-V characteristics and performance parameters of ternary solar cells of the polymer P3HT blended with stoichiometric fullerene mixtures. The total P3HT:fullerene ratio and the total fullerene content remain constant. X represents the weight percent of $PC_{70}BM$ and, correspondingly 100-X is the weight percentage of $PC_{60}BM$ or $IC_{60}BA$. The J-V curves are averaged for several (5 to 15) properly working devices to better highlight the subtle changes, especially for the PCBM-based devices. In both cases, the short circuit current $J_{SC}$ increases for an increasing amount of $PC_{70}BM$. This $J_{SC}$ increase as a function of the weight percent of $PC_{70}BM$ is explained by its higher optical absorption in comparison to both, $PC_{60}BM$ and $IC_{60}BA$.

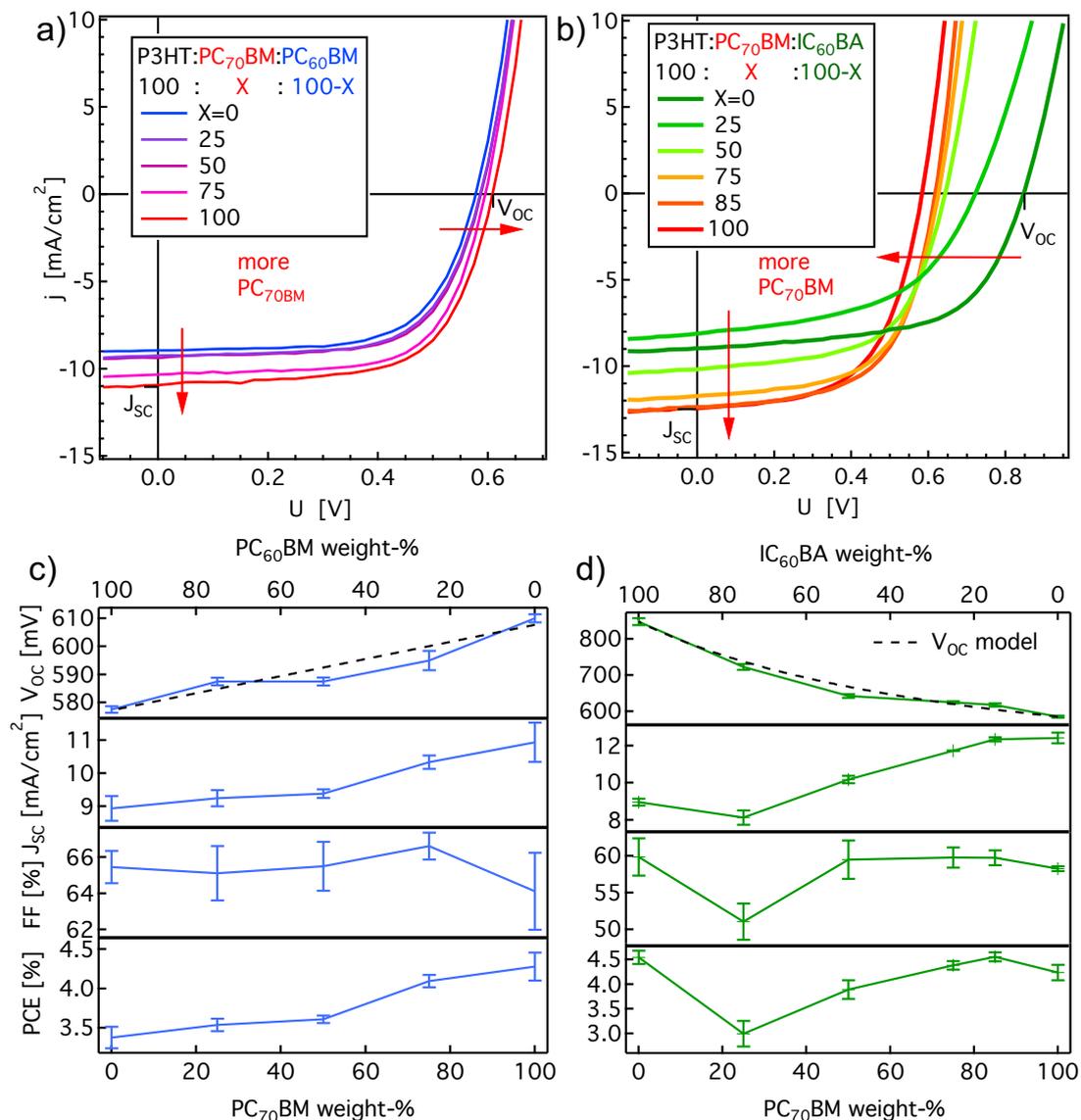

**Figure 2**. Ternary solar cell performance of P3HT blended with **a,c)** $PC_{70}BM:PC_{60}BM$ and **b,d)** $PC_{70}BM:IC_{60}BA$. The total P3HT:fullerene ratio remains constant. X represents the weight percent of $PC_{70}BM$ with respect to the second fullerene. In both cases, increasing the amount of $PC_{70}BM$ continuously increases $J_{SC}$ due its stronger optical absorption. In **a,c)** the open circuit voltage $V_{OC}$ follows the $PC_{70}BM$ weight percent linearly (black dashed line), leading to an increase of 30 mV. In **b,d)** $V_{OC}$ decreases drastically with increasing $PC_{70}BM$ percentage. The $V_{OC}$ model (black dashed line) will be discussed later. The error bars in **c,d)** represent the standard deviation for 5 to 15 solar cells.

For P3HT:PC$_{70}$BM:PC$_{60}$BM $J_{SC}$ increases from 9 mA/cm$^2$ to 11 mA/cm$^2$ for 0 – 100 weight-% of PC$_{70}$BM. Additionally, $V_{OC}$ increases by 30 mV from 580 mV to 610 mV. This is a surprising result, as the PC$_{70}$BM LUMO is assumed to lie around 6 – 10 meV lower in energy than the PC$_{60}$BM LUMO as shown in **Figure 1**.[11,20] An explanation might be the inhibited aggregation of PC$_{60}$BM upon adding PC$_{70}$BM. This increases the charge transfer state energy ($E_{CT}$) of P3HT:PC$_{60}$BM, leading to higher $E_{CT}$ and $V_{OC}$.[27] On the other hand, this effect might also be caused by the different density of state (DOS) distributions of the fullerene derivatives, as they are close in energy and the DOS may overlap.[6] As the DOS of fullerenes is only a few tens of meV wide, this assumption holds true for at least PC$_{60}$BM:PC$_{70}$BM based blends.

For P3HT:PC$_{70}$BM:IC$_{60}$BA $J_{SC}$ increases from 9 mA/cm$^2$ to 13 mA/cm$^2$ for 0 – 100 weight-% of PC$_{70}$BM. In contrast to the previous case, $V_{OC}$ decreases tremendously by 220 mV from 820 mV to 600 mV upon increasing the PC$_{70}$BM content from 0% to 100%. The PC$_{70}$BM LUMO lies around 170 – 210 meV lower in energy than the IC$_{60}$BA LUMO as depicted in **Figure 1**. The $V_{OC}$ decrease is thus a direct consequence of the LUMO level difference. Remarkably, the mixture with 75:25 IC$_{60}$BA:PC$_{70}$BM exhibits the lowest short circuit current. We therefore assume PC$_{70}$BM to act as a current-limiting electron trap for this stoichiometry, due to the lower LUMO level in comparison to IC$_{60}$BA.

## 4. Charge Carrier Distribution in Ternary Blends
### 4.1 EPR Signatures of Light-Induced Charge Carriers

Before we can use EPR to analyse the distribution of charge carriers between the donor and acceptor molecules and quantify their contributions, we have to identify the exact spectral signatures involved and perform control experiments to ensure that the extracted charge carrier densities are not distorted by the choice of experimental conditions.

**Figure 3a** shows light-induced EPR spectra of the polymer P3HT blended with the three different fullerene derivatives. The observed lineshapes are similar for all three samples and consist of contributions from the radical cation P3HT$^+$ and the respective fullerene anions PC$_{70}$BM$^-$, PC$_{60}$BM$^-$ and IC$_{60}$BA$^-$. All these signals have different $g$-factors, which results in different positions along the magnetic field axis. The experimental spectra are superpositions of the individual components, but they can be deconvoluted to individual components by simulations with the MatLab tool EasySpin.[28] Important to note, the intensity of each component depends on the absolute number of charge carriers (spins) present in the sample. The simulation parameters are summarized in **Table 1**.

**Table 1.** $g$-tensors for EasySpin simulations.

| $g$ | PC$_{70}$BM$^-$ | PC$_{60}$BM$^-$ | IC$_{60}$BA$^-$ | P3HT$^+$ | PTB7$^+$ | PCDTBT$^+$ |
|---|---|---|---|---|---|---|
| $g_{xx}$ | 2.00543 | 2.00052 | 2.00075 | 2.00337 | 2.0045 | 2.0032 |
| $g_{yy}$ | 2.00265 | 2.00031 | 2.00023 | 2.00232 | 2.0031 | 2.0024 |
| $g_{zz}$ | 2.00235 | 1.99931 | 2.00295 | 2.00107 | 2.0023 | 2.0018 |

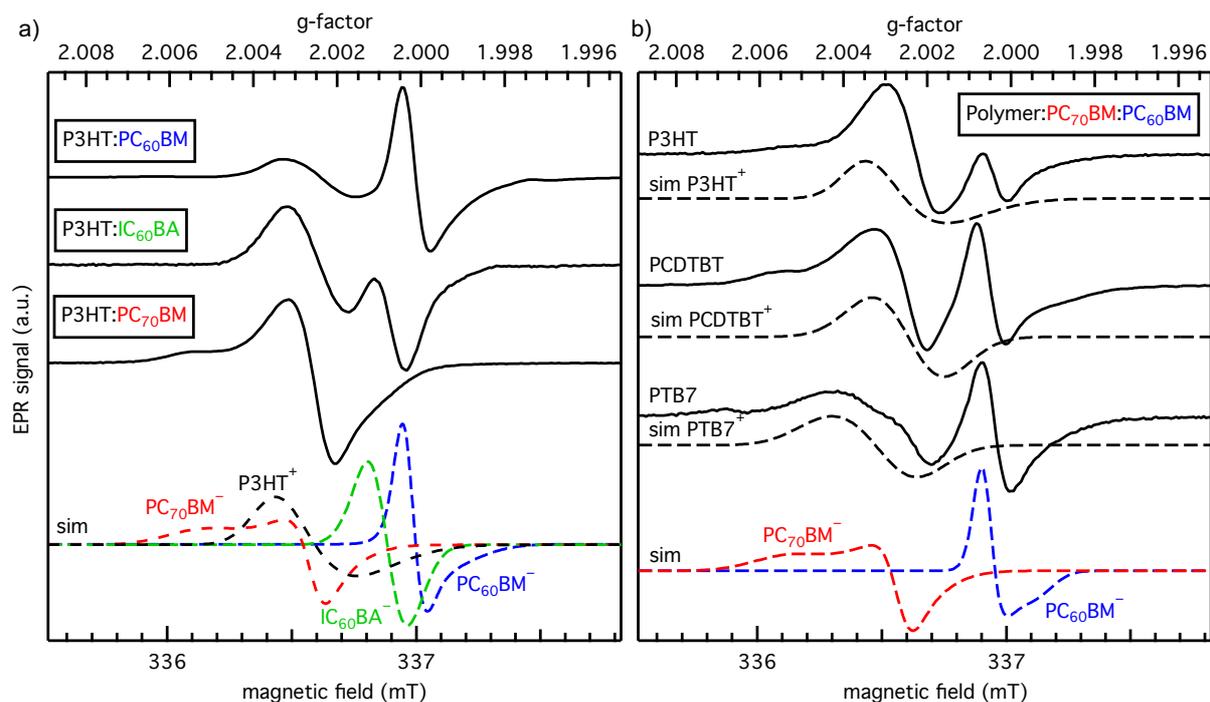

**Figure 3**. Light-induced EPR spectra (solid black lines) of polymer:fullerene blends (*T* = 30 K) together with simulations (dashed lines) of all contributions. **a)** P3HT blended with three different fullerene derivatives. **b)** P3HT, PCDTBT and PTB7 blended with a 1:1 mixture of PC$_{70}$BM:PC$_{60}$BM. The detected spectra are superpositions of the respective individual anion and cation signatures.

To also test, whether the choice of the polymer has a relevant influence, we performed similar measurements with blends of the three polymers P3HT, PCDTBT and PTB7 with a 1:1 mixture of PC$_{70}$BM:PC$_{60}$BM as shown in **Figure 3b**. As in the previous samples, the experimental spectra are superpositions of the individual anion and cation contributions. In all spectra shown in **Figure 3b**, we can clearly see contributions of both fullerene radical anions, PC$_{70}$BM$^-$ and PC$_{60}$BM$^-$. This does already indicate that electrons are not predominantly localized on one type of fullerene. Furthermore, there is no indication that the distribution of electrons between the two fullerenes is influenced by the choice of the donor polymer.

**4.2 Influence of Temperature, Excitation Wavelength and Thermal Annealing**
EPR spectra were recorded at temperatures of 100 K and below to achieve sufficient signal-to-noise ratio, needed to perform a quantitative analysis of the respective fullerene contributions. Overall EPR signal intensities were roughly the same for the changing blend ratios at each temperature. One might have expected additional charge carrier trapping at low temperatures, because of a larger number of deep traps in PC$_{70}$BM relative to PC$_{60}$BM.[14,15] We can infer that this does not seem to be a relevant effect for these ternary blend systems: mixing fullerenes does not result in additional charge trapping.

Additionally, we tested for a possible dependence of the EPR spectra on the excitation wavelength. Since the fullerene derivatives have different absorption spectra, they absorb different amounts of the incident light in a solar cell. This leads to photoinduced hole transfer from fullerene acceptors to the polymer donor. This in turn would yield an uneven photogenerated electron density. We used a halogen bulb and a LED as white light sources and a 532 nm laser at various illumination intensities. Additionally, in order to check for the effect of morphology, illumination dependent EPR spectra of the PCBM-based blends were recorded before and after thermal annealing (10 min at 130°C). The resulting spectra differ in overall signal intensity, but they are perfectly congruent after normalization. Hence, neither the morphology, nor illumination intensity or wavelength have any effect on the distribution of electrons between the fullerene acceptors.

Furthermore, we would like to point out that the shown EPR spectra reflect steady state populations after charge transfer and relaxation of the charge carriers in the blend films. Taking this all into account, we assume that the distribution of electrons between involved fullerene derivatives depends solely on their energetics and respective stoichiometry.

Having established that EPR is a suitable method to detect and distinguish spectroscopic signatures of anions and cations in ternary solar cell blends, we use this approach in the following to determine the electron distribution between different fullerenes.

**4.3 Quantifying Electron Distributions Between Fullerene Derivatives**

To quantify the electron distribution between fullerene derivatives, we prepared two sample series of P3HT blended with stoichiometric fullerene mixing ratios. **Figure 4** shows the corresponding EPR spectra of P3HT:$PC_{70}BM$:$PC_{60}BM$ and P3HT:$PC_{70}BM$:$IC_{60}BA$ with blending ratios given as 100:*X*:100−*X*. The gradual change of the spectra from 0% (no $PC_{70}BM$) to 100% (only $PC_{70}BM$) is reflected by the shifting intensities between the contributions of $PC_{70}BM^-$, $PC_{60}BM^-$ and $IC_{70}BA^-$ to the spectrum, as indicated by the vertical arrows. While this stepwise intensity shift is clearly visible for P3HT:$PC_{70}BM$:$PC_{60}BM$ in **Figure 4a**, for P3HT:$PC_{70}BM$:$IC_{60}BA$ in **Figure 4b** a rapid decrease in $IC_{60}BA^-$ intensity is observed when adding just small amounts (2 − 5%) of $PC_{70}BM$. This already indicated different energetics for electron localization among available fullerene acceptors. For the following quantitative analysis, the spectra were fitted as superpositions of the spectral components of anions and cations.

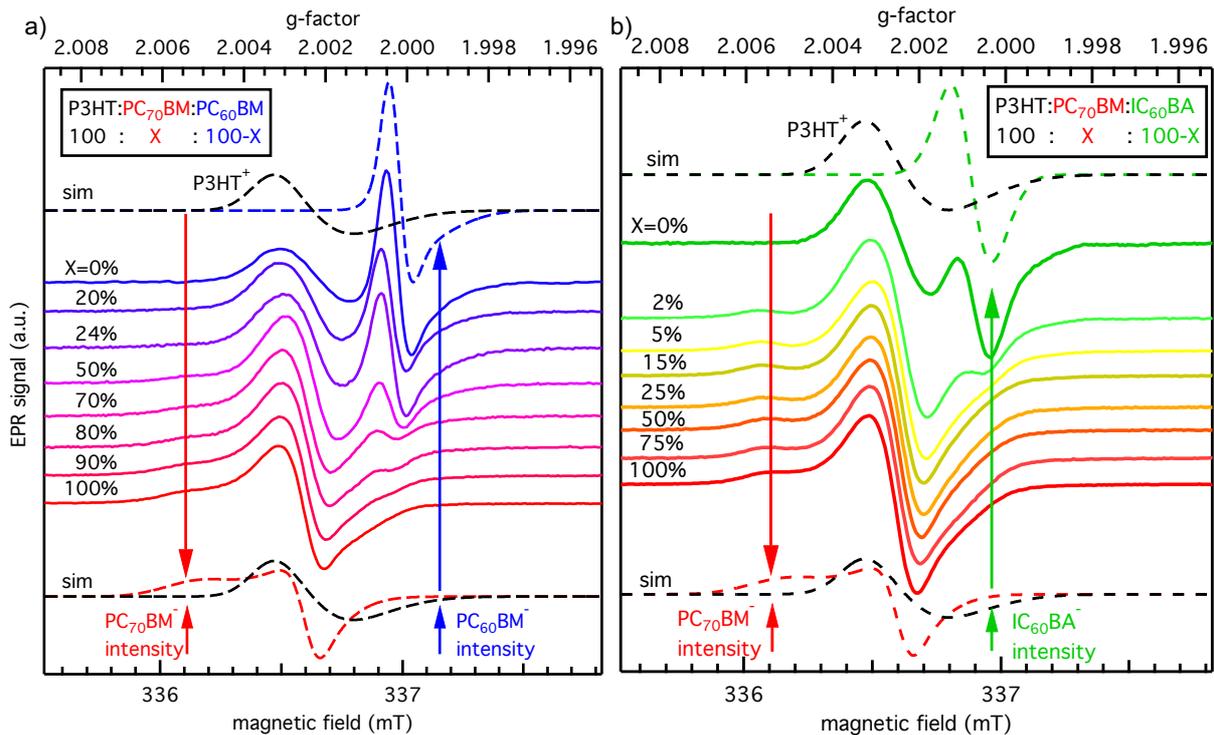

**Figure 4.** Light-induced EPR spectra (solid lines) of ternary blends with stoichiometric fullerene ratios ($T$ = 30 K) together with simulations (dashed lines) of all contributions. **a)** For P3HT:$PC_{70}BM$:$PC_{60}BM$ a gradual redistribution of signal intensity between the two fullerene anion contributions can be observed for changing stoichiometry. **b)** For P3HT:$PC_{70}BM$:$IC_{60}BA$ a rapid decrease in $IC_{60}BA^-$ intensity is observed when adding small amounts (2 – 5%) of $PC_{70}BM$. The vertical arrows indicate the shifting intensities between the respective fullerene derivatives.

The fitted intensities are proportional to the respective number of electrons localized on either fullerene $N_{60}$ or $N_{70}$. Thereby, their percentage $P$ can be expressed as $P_{70}(X) = N_{70}(X)/(N_{70}(X) + N_{60}(X))$, with $P_{60}(X) + P_{70}(X) = 100\%$, which resembles normalized relative intensities of the spectral components. **Figure 5** shows the resulting electron distributions as a function of the fullerene weight ratio $X$. Error bars in intensity are due to the low signal to noise ratios and overlapping signals resulting in fitting uncertainties. Errors in the weight ratios due to the uncertainties in weighing are negligible. For P3HT:$PC_{70}BM$:$PC_{60}BM$ in **Figure 5a** an almost linear dependence between the fullerene weight percentage and the percentage of electrons localized on the respective fullerene derivative is observed. For comparison, diagonal dotted lines are added (red), which represent an even distribution according to $X$. One can clearly see that within the error bars, such an ideal distribution is already a good approximation of the experimental values. In contrast to that, for P3HT:$PC_{70}BM$:$IC_{60}BA$ in **Figure 5b** electrons show a strong preference to localize on $PC_{70}BM$ when just adding a few weight percent.

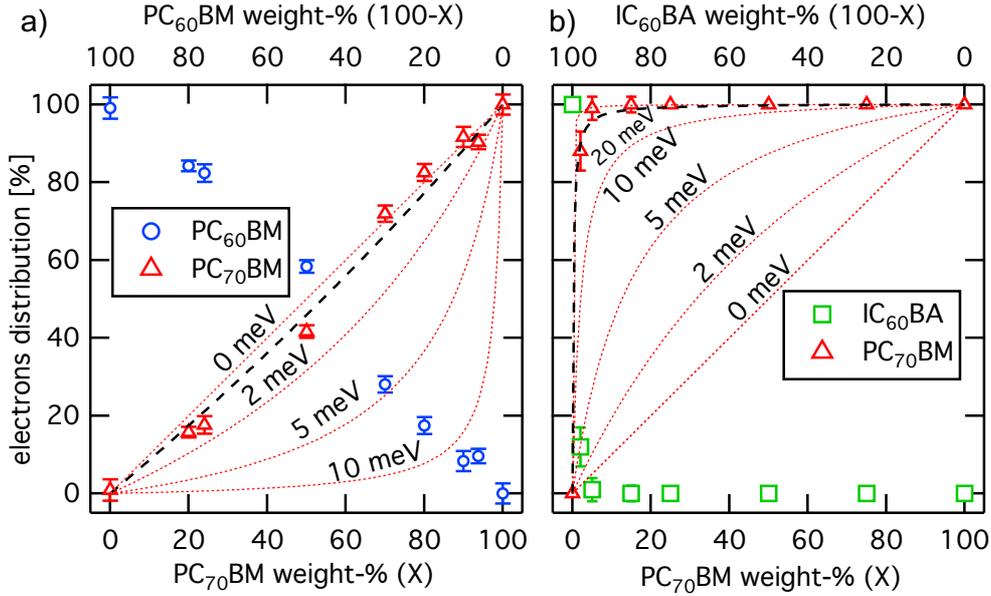

**Figure 5.** Electron distribution on **a)** PC$_{70}$BM:PC$_{60}$BM and **b)** PC$_{70}$BM:IC$_{60}$BA as a function of the fullerene weight ratio $X$ ($T$ = 30 K). The electron distribution between PC$_{70}$BM:PC$_{60}$BM closely resembles the weight ratio $X$, while for PC$_{70}$BM:IC$_{60}$BA electrons predominantly localize on PC$_{70}$BM. This is caused by the energetic difference an electron experiences for localizing on either fullerene derivative. The black dashed lines are fits using the Boltzmann distribution from **Equation 1**. Dotted red curves are exemplary distributions for the energy of an electron being **a)** a few meV higher on PC$_{70}$BM than on PC$_{60}$BM, **b)** a few meV lower on PC$_{70}$BM than on IC$_{60}$BA.

Next, to estimate the energetic ΔLUMO difference for an electron being localized on either fullerene, we approximate the electron populations ($N_{60}$, $N_{70}$) by a two-level system separated by an energy gap ΔLUMO. The population of such a system can be described by a Boltzmann distribution:

$$\frac{N_{60}}{N_{70}} = \frac{P_{60}(X)}{P_{70}(X)} = \frac{X}{1-X} exp\left(\frac{-\Delta\text{LUMO}}{k_B T}\right) \quad (1)$$

Here, $T$ is the temperature and $k_B$ the Boltzmann constant. The fraction $X/(1-X)$ corresponds to the statistical weight of the two energy levels, given by the stoichiometric fullerene weight ratio $X$. In order to verify the energy resolution, exemplary Boltzmann distributions are added to **Figure 5** (dotted red curves). They illustrate the energy of an electron on PC$_{70}$BM being on the same level (0 meV) and 2, 5, 10 or even 20 meV higher (**Figure 5a**) or lower (**Figure 5b**) than on the respective second fullerene derivative. For only a few meV ΔLUMO energy difference, this results in a strong occupation preference of the lower-lying fullerene. We fitted this distribution to the two data sets in **Figure 5**, resulting in the curved dashed lines (black) and values for ΔLUMO of (0 ± 1) meV for PC$_{70}$BM:PC$_{60}$BM and (20 ± 5) meV for PC$_{70}$BM:IC$_{60}$BA. These values are approximately just one tenth of the expected LUMO differences as discussed for **Figure 1**, where (6 – 10) meV and (170 – 210) meV are reported for the pristine materials. We could confirm these low ΔLUMO values with additional exemplary EPR measurements for annealed and not annealed samples and for temperatures of 10, 20, 70 and 100 K. This excludes sample treatment or experimental temperature as relevant factors for the ΔLUMO determination. The very low ΔLUMO values indicate strong interaction between blended fullerenes and LUMO hybridisation resulting in the formation of an effective LUMO level.

## 5. Discussion

Now we compare the two estimated ΔLUMO values from the EPR analysis with the respective shifts in $V_{OC}$ as shown in **Figure 2**. For both ternary blend systems, a shift in $V_{OC}$ is observed when changing the fullerene stoichiometry. In the case of P3HT:PC$_{70}$BM:PC$_{60}$BM a linear trend is observed (**Figure 2c**), while a curved trend can be modelled for P3HT:PC$_{70}$BM:IC$_{60}$BA (**Figure 2d**). This curved trend is the result of scaling **Equation 1** in between the starting ($X$ = 0) and end ($X$ = 100) voltages for $T$ = 300 K using the ΔLUMO difference of 20 meV that has been determined from the EPR analysis. The same is true for the linear trend in **Figure 2c** for ΔLUMO = 0 meV. Remarkably, the $V_{OC}$ trends resemble the same ΔLUMO values, which proves that the blended fullerenes can actually be described by an effective LUMO level.

The different starting and end voltages for P3HT:PC$_{70}$BM:IC$_{60}$BA can be directly explained by the different fullerene LUMO levels. On the other hand, the different voltages for P3HT:PC$_{70}$BM:PC$_{60}$BM can be caused by the inhibited aggregation of PC$_{60}$BM upon adding PC$_{70}$BM, which increases the CT state energy and hence $V_{OC}$.[27] However, in both cases, $V_{OC}$ scales for the intermedium stoichiometries according to a weighted two-state Boltzmann distribution with a 10-fold decrease in ΔLUMO.

In comparison to studies investigating the performance of similar ternary OPV systems, consisting of one polymer blended with two different C$_{60}$-derivative acceptors (PC$_{60}$BM, IC$_{60}$BA) [29,30] or in general a broader variety of ternary OPV systems,[31] it seems that our results are best described by an alloying effect of the two fullerene derivatives. This fullerene alloy is then characterized by an averaged, effective LUMO level, while the individual molecules maintain their characteristic optical absorption. Due to its higher optical absorption coefficient, one can indeed expect PC$_{70}$BM to generate more initial excitons than the C$_{60}$-derivatives. However, this initial excitation imbalance seems to vanish as the diffusion towards the heterojunction and the subsequent charge carrier separation and relaxation take place.

## 6. Conclusion

Motivated to understand the correlations between the energetics of acceptor (A) molecules and the efficiency of electron transfer processes from photoexcited donors (D) and their impact on the open circuit voltage ($V_{OC}$) of ternary solar cells (D:A:A), we investigated combinations of three different donor polymers blended with three different fullerenes as a function of the fullerene stoichiometry. Using quantitative electron paramagnetic resonance (EPR), we were able to identify and distinguish spectroscopic signatures of light-induced polymer cations and fullerene anions in the ternary blends. Careful analysis of the EPR spectra yielded the relative distribution of electrons between the respective fullerene acceptors. We found that this distribution is not only governed by the stoichiometry, but also by the effective LUMO difference of the studied fullerene-derivatives. Within a two-level model and assuming Boltzmann distribution, we estimated the LUMO difference to be in the order of (20 ± 5) meV for PC$_{70}$BM:IC$_{60}$BA and less than 1 meV for PC$_{70}$BM:PC$_{60}$BM acceptor blends, i.e., a tenth of the LUMO difference reported for separately measured pristine materials. Remarkably, the same Boltzmann distribution scaled to the $V_{OC}$ data can fully reproduce the observed dependence of $V_{OC}$ on the fullerene stoichiometry in ternary solar cells. This shows that mixed (or unpurified) C$_{60}$- and C$_{70}$-based fullerene derivatives, despite their differences when studied separately, can be described by an effective LUMO level in a mixed phase, e.g., due to the delocalization of an electron wave function over two molecules.

**Acknowledgements**
We thank Dr. Stefan Väth for preliminary EPR measurements. This work was supported by the Deutsche Forschungsgemeinschaft (DFG, German Research Foundation) within SPP1601 (DY18/11-1 and SP1563/1-1) and the Research Training School ''Molecular biradicals: Structure, properties and reactivity" (GRK 2112). Further, we acknowledge support by the Bavarian Network "Solar Technologies Go Hybrid" (SolTech). Open Access funding enabled and organized by Projekt DEAL.